\documentclass[a4paper,11pt]{article}
\usepackage{jcappub} 
\usepackage{lineno}
\usepackage{longtable}
\usepackage{supertabular}
\usepackage{booktabs}
\usepackage{multirow}
\usepackage[utf8]{inputenc}   
\DeclareUnicodeCharacter{2212}{-}

\arxivnumber{} 
\title{\boldmath CLASH-VLT: Constraining deviation from GR with the mass profiles of nine massive galaxy clusters}

\author[a,b]{L. Pizzuti,}
\author[b,c]{A. Biviano,}
\author[d]{K. Umetsu,}
\author[a]{E. Agostoni,}
\author[a]{A. Autorino,}
\author[e]{A. M. Pombo,}
\author[f,g,h]{A. Mercurio}
\author[g]{and M. D'Addona}
\affiliation[a]{Dipartimento di Fisica G. Occhialini, Universit\`a di Milano-Bicocca, \\
Piazza della Scienza 3, 20126 Milano, Italy}

\affiliation[b]{INAF-Osservatorio Astronomico di Trieste, via G. B. Tiepolo 11, 34143 Trieste, Italy}
\affiliation[c]{IFPU-Institute for Fundamental Physics of the Universe, via Beirut 2, 34014 Trieste, Italy}
\affiliation[d]{Academia Sinica Institute of Astronomy and Astrophysics (ASIAA), No. 1, Section 4, Roosevelt Road, Taipei 106216, Taiwan}
\affiliation[e]{CEICO - Institute of Physics of the Czech Academy of Sciences}
\affiliation[f]{Università di Salerno, Dipartimento di Fisica “E.R. Caianiello”, Via Giovanni Paolo II 132, 84084 Fisciano (SA), Italy.}
\affiliation[g]{INAF Osservatorio Astronomico di Capodimonte, Salita Moiariello 16, 80131 Napoli, Italy.}  
\affiliation[h]{INFN–Gruppo Collegato di Salerno–Sezione di Napoli, Dipartimento di Fisica “E.R. Caianiello”, Università di Salerno, Via Giovanni Paolo II, 132, 84084 Fisciano (SA), Italy.}

\emailAdd{lorenzo.pizzuti@unimib.it}

\abstract{We investigate the anisotropic stress parameter, $\eta=\Psi/\Phi$, defined as the ratio of the gravitational potentials in the linearly perturbed Friedmann-Lemaître Robertson-Walker metric, as a probe of deviations from general relativity across astrophysical to cosmological scales. Using mass profiles reconstructed from high-precision lensing and kinematics of nine galaxy clusters from the CLASH-VLT sample, we derive $\eta(r)$ as a function of the radial distance from the cluster centres, over the range $[0.1 \,\text{Mpc},1.2\,r_{200}^L]$, where $r_{200}^L$ is virial radius best-fit from lensing data. When using a Navarro-Frenk-White or an Hernquist profile to model the total matter distribution, we find consistency with general relativity ($\eta = 1$) within $2\sigma$ for the full radial range for all the sampled clusters. However, adopting a Burkert profile introduces mild tension with general relativity, reaching the $3\sigma$ level in two systems. Assuming a negligible time-dependence in the redshift range spawned by the clusters, we obtain the joint constraint $\eta (r= 1.0 \, \text{Mpc}) = 0.93^{+0.48}_{-0.40}$ (stat) $\pm 0.47$ (syst) at $95\% $ confidence level -- an improvement of approximately $40\%$ over previous estimates. We discuss the impact of systematics on the constraints, and we highlight the implications of this result for current and upcoming cluster surveys.
}

\begin{document}
\maketitle
\flushbottom
%
\section{Introduction}\label{sec:intro}
%
    The discovery of the Universe's accelerated expansion (\textit{e.g.} Ref. \cite{Riess:1998cb,Perlmutter:1998np}), more than two decades ago, remains one of the most profound mysteries in modern cosmology. The current best description of a large variety of independent cosmological phenomena across different scales (\textit{e.g.} Ref. \cite{article}) is the $\Lambda$CDM model. In this scenario, the observed acceleration is attributed to a cosmological constant, $\Lambda$, present in the field equations of Einstein's General Relativity (GR). However, the physical origin of this constant remains unknown (\textit{e.g.} Ref. \cite{Padilla15, fadecollaboration2022}).

    Recent observations on both large and small scales have increasingly highlighted tensions between the predictions of the $\Lambda$CDM model and empirical data, as well as discrepancies among independent datasets. Notable examples include the $S_8$ tension (\textit{e.g.} Ref. \cite{Abbott_2022, PhysRevD.110.123508}), the CMB anisotropy anomalies (\textit{e.g.} Ref. \cite{Schwarz_2016, Hansen_2023}), the $5 \sigma$ Hubble tension (\textit{e.g.} Ref. \cite{hubble_tension2023}) between the high-redshift estimates of $H_0$ from the Cosmic Microwave Background (CMB) data (\textit{e.g.} Ref. \cite{Di_Valentino_2021}) and local measurements based on distance indicators (\textit{e.g.} Ref. \cite{Riess_2022}). Additionally, recent analyses combining DESI and CMB data indicate a time-dependent equation of state for Dark Energy (DE) (\textit{e.g.} Ref. \cite{Adame_2025}).

    However, whether these tensions arise from an unknown systematic effect or signal the need for new physics beyond the $\Lambda$CDM model remains an open question. Nevertheless, the increasingly complex observational landscape has motivated the exploration of alternative models to the standard cosmological framework. Over the last 20 years, many theoretical models have been proposed to address these tensions. In addition to addressing these discrepancies, such models often seek to offer a more fundamental understanding of the nature of dark matter (DM) and DE components -- introduced in $\Lambda$CDM to account for observations -- which together constitute approximately 95\% of the Universe's total energy/matter budget.

    A possible solution is to extend the $\Lambda$CDM model by introducing new interactions within the dark sector (DE+DM) or incorporating exotic fields characterized by non-standard equations of state. Among the various proposed possibilities, particular interest has been given to Modified Gravity (MG) theories (\textit{e.g.} Ref. \cite{Koyama_2016}), in which GR is modified to mimic the effect of the dark sector at large scales in order to reproduce observations.

    Whether by modifying gravity or redefining the physics of the dark sector, viable extensions to the $\Lambda$CDM model inevitably introduce new degrees of freedom that influence the formation and evolution of cosmic structures.
    Galaxy clusters - the largest structures in the Universe that have undergone gravitational collapse -- are an excellent target of cosmological studies. Most of the matter content in clusters (roughly 85\%) is understood to be in the form of invisible collisionless DM (\textit{e.g.} Ref. \cite{Biviano06, Sartoris2020}). While processes involving ordinary (baryonic) matter play a central role in the core of these structures, the overall cluster's dynamics is dominated by the gravitational interaction, making them ideal laboratories for investigating deviations from GR and probing the properties of the dark sector. Their internal mass distribution is dependent on the underlying cosmology, and they further probe both the relativistic and non-relativistic sectors of the gravitational interaction, through lensing and internal kinematics (of gas or member galaxies), respectively. For this purpose, galaxy clusters have been extensively used to test MG models through their mass profile reconstruction (with kinematics or lensing techniques, \textit{e.g.} Ref. \cite{Pizzuti_2016}), abundance and spatial distribution (\textit{e.g.} Ref. \cite{Vogt2024, howard2022}).
    
    One way to parametrize generic departures from standard gravity is to study the anisotropic stress $\eta$ -- \textit{aka} gravitational slip. $\eta$ is defined as the ratio between the relativistic and non-relativistic gravitational potentials, $\Psi/\Phi$, that are components of the linearly perturbed Friedmann-Lemaitre-Robertson-Walker (FLRW) metric. In GR, and in models where additional degrees of freedom are minimally coupled to gravity, $\eta = 1$; this is no longer true in a general MG scenario. Here and in the following we further assume that DM is pressureless, and its contribution to the anisotropic stress tensor is negligible.  Thus, a genuine departure from $\eta=1$ can be linked to a departure from GR (\textit{e.g.} Ref. \cite{anton2025gravitationalslipparameterizedpostnewtonian}).

    In this work, we constrain $\eta$ by using precise determinations of cluster mass profiles obtained from gravitational lensing and kinematic analyses of member galaxies. While galaxies move under the influence of the non-relativistic potential, $\Phi$, photons propagate along null geodesics, reflecting the contribution of both relativistic, $\Psi$, and non-relativistic potentials, $\Phi_\text{lens} = (\Psi + \Phi)/2$. A proper combination of lensing and kinematics mass estimates of clusters can thus be used to probe deviations from GR, under the assumption that possible systematic effects are under control.

    Through a similar approach to the one in Ref.~\cite{Pizzuti_2016}, we compute the radial profile $\eta(r)$ in the region $r> 0.1 \,\text{Mpc}$ for nine clusters. These nine clusters were chosen based on the availability of high-precision imaging and spectroscopic data obtained from the Cluster Lensing and Supernova Survey with Hubble (CLASH) collaboration, along with its follow-up observations using the VIMOS spectrograph at the Very Large Telescope (CLASH-VLT). The kinematic mass profiles were reconstructed by analysing the positions and line-of-sight velocities of cluster members with the \textsc{MG-MAMPOSSt} code (\textit{e.g.} Ref.~\cite{Pizzuti2021}). We then combined these results with the strong+weak gravitational lensing posteriors from Ref.~\cite{Umetsu16}. In our analysis, we explore various models for both lensing and kinematic mass profiles, while also addressing the impact of systematics. A particular focus is placed on deviations from spherical symmetry and the equilibrium configuration.

    The paper is structured as follows: Section \ref{sec:theo} reviews the definition of the anisotropic stress and its connection to the clusters' total mass profiles; in Section \ref{sec:data}, we present the dataset, further elaborating on the kinematic mass reconstruction with \textsc{MG-MAMPOSSt}. Section \ref{sec:results} is devoted to the main results, which are further summarised in Section \ref{sec:conc}.

    Throughout this paper, we assume a flat $\Lambda$CDM cosmology with $H_0 = 70$ km s$^{-1}$ $\text{Mpc}^{-1}$, $h = H_0/(100$ km s$^{-1}$ $\text{Mpc}^{-1}$), $\Omega_m = 0.3$, $\Omega_\Lambda = 0.7$.
%
 \section{Theoretical Background}\label{sec:theo}
%
    While galaxy clusters are highly non-linear perturbations of the matter density field, their geometry can still be described by a linear perturbation of the FLRW metric (\textit{e.g.} Ref.~\cite{Pizzuti_2016}). In the conformal Newtonian gauge, the line element reads:
    \begin{equation}
     \text{d}s^2 = -\left(1+\frac{2\,\Phi}{c^2}\right)c^2\text{d}t^2 + a^2(t)\left(1 -\frac{2\,\Psi}{c^2}\right)\left[\text{d}r^2 + r^2\text{d}\Omega^2\right]\,,
    \end{equation}
    where $\Phi(t,\vec{x})$ and $\Psi(t,\vec{x})$ are the two gauge-invariant non-relativistic and relativistic potentials, respectively. In GR, for a fluid with a negligible anisotropic stress-energy tensor, one has $\Phi = \Psi$ at all points in spacetime, a relation valid up to second-order corrections ($\ll 1$) in the relativistic expansion of the potentials. Assuming such a configuration, generic departures from standard gravity can be quantified with the anisotropic stress $\eta =\Psi/\Phi$, which is a general function of scale and time and whose specific form depends on the assumed model. Following the work done in Ref.~\cite{Pizzuti_2016,Pizzuti19}, we employ a phenomenological, nearly model-independent determination of $\eta(r,t)$ using the mass profiles of galaxy clusters derived from kinematic and lensing analysis. The dependence on the scale is encapsulated in the free parameters of the mass density profiles.

    Under the assumption of dynamical relaxation, member galaxies in clusters move smoothly under the non-relativistic potential, $\Phi$, and follow the Jeans' equation:
    \begin{equation}\label{eq:jeans}
     \frac{\text{d} (\nu \sigma_r^2)}{\text{d} r}+2\beta(r)\frac{\nu\,\sigma^2_r}{r}=-\nu(r)\,\frac{\text{d} \Phi}{\text{d} r}\,,
    \end{equation}
    where we have further assumed spherical symmetry; $\nu(r)$, is the number density profile of tracers, as a function of the 3-dimensional radial distance from the cluster center; $\sigma^2_r$, is the velocity dispersion along the radial direction, and $\beta \equiv 1-(\sigma_{\theta}^2+\sigma^2_{\varphi})/2\sigma^2_r$ is the so called velocity anisotropy. The $\sigma_{\theta}^2$ and $\sigma^2_{\varphi}$ are the velocity dispersion components along the tangential and azimuthal directions, respectively. Spherical symmetry imposes $\sigma_{\theta}^2=\sigma^2_{\varphi}$ and the expression of the anisotropy profile simplifies to $\beta = 1 -\sigma_{\theta}^2/\sigma_{r}^2$.

    The gradient of the potential is connected to the total (effective) dynamical mass through the Poisson equation,
    \begin{equation}\label{eq:dyn}
     \frac{\text{d} \Phi}{\text{d} r} = \frac{GM_\text{dyn}(r)}{r^2}\,,
    \end{equation}
    where 
    \begin{equation}
     M_\text{dyn}(r) = 4\pi \int_0^r{r'^2\, \rho_\text{dyn}(r')\,\text{d}r'}\,,
    \end{equation}
    with $\rho_\text{dyn}(r)$ the effective dynamical density\footnote{The term "effective" indicates that $\rho_\text{dyn}(r)$ may include contributions coming from additional degrees of freedom in theories beyond $\Lambda$CDM, which are not explicitly modeled.}. 
    
    Concerning the lensing mass, we can derive a similar expression by considering that photon propagation is affected by both relativistic and non-relativistic potentials. The potential experienced by photons — the source of gravitational lensing, commonly referred to as the Weyl potential (\emph{e.g.} Ref. \cite{Tutusaus_2024}) — is defined as $\Phi_\text{lens} =(\Phi +\Psi )/2$, which satisfies the analogous of Eq.~\eqref{eq:dyn} for an effective lensing mass,
    \begin{equation}\label{eq:lens}
     \frac{\text{d} (\Phi+\Psi)}{\text{d} r} = 2\,\frac{GM_\text{lens}(r)}{r^2}\,.
    \end{equation}

    The resulting anisotropic stress comes as 
    \begin{equation}\label{eq:eta}
     \eta(r) = \frac{\int_{r}^\infty{\Big[ 2M_\text{lens}(r')-M_\text{dyn}(r')\Big]\frac{\text{d}r'}{r'^2}}}{\int_{r}^\infty{M_\text{dyn}(r')\frac{\text{d}r'}{r'^2}}}\,.
    \end{equation}
%

%
    \subsection{Mass modeling}
%
     In this work, we adopt three different models to describe the total effective mass profiles of the clusters under analysis. The models were chosen such that they provide adequate fits for both kinematics and lensing data (see Section \ref{sec:data}). All models are characterized by two free parameters: the "virial" radius $r_{200}$, which is the radius of a sphere enclosing an average density 200 times the critical density of the Universe at the cluster's redshift; and a scale radius $r_\text{s}$, characterizing the change of slope from the inner to the outer part of the matter distribution. The first model to consider is the popular Navarro-Frenk-White (NFW) profile (\textit{e.g.} Ref.~\cite{Navarro:1995iw}) is given by
        \begin{equation}
         M_\text{NFW}(r) = M_{200}\frac{\text{ln}(1+r/r_\text{s}) - \frac{r/r_\text{s}}{(1+r/r_\text{s} )}  }{\text{ln}(1+r_{200}/r_\text{s}) - \frac{r_{200}/r_\text{s}}{(1+r_{200}/r_\text{s} )}  }\,,
        \end{equation}
    where $r_\text{s} \equiv r_{-2}$ is the radius at which the logarithmic derivative of the density profile is $-2$.

    The second is the Burkert model (\textit{e.g.} Ref.~\cite{Burkert01}), defined as
        \begin{equation}
         M_\text{Bur}(r) = M_{200}\frac{\text{ln}\big[1+(r/r_\text{s})^2\big] +  2\,\text{ln}(1+r/r_\text{s}) - 2\,\text{arctan}(r/r_\text{s}) }{\text{ln}\big[1+(r_{200}/r_\text{s})^2\big] +  2\,\text{ln}(1+r_{200}/r_\text{s}) - 2\,\text{arctan}(r_{200}/r_\text{s})}\,,
        \end{equation}
    where $r_\text{s} \simeq 2/3 \, r_{-2}$. Finally, we consider the Hernquist model profile (\textit{e.g.} Ref.~\cite{Hernquist01}),
        \begin{equation}
         M_\text{Her}(r) = M_{200}\frac{(r_{200}+ r_\text{s})^2}{r_\text{200}^2}\frac{ r^2}{(r+ r_\text{s})^2}\,,
        \end{equation}
    with $r_\text{s} =2\, r_{-2}$. Note that, while the NFW and Hernquist models have a central density which diverges as $r^{-1}$, the Burkert model exhibits a core. For all three models, analytical expressions of $\eta(r)$ can be derived by integrating the total mass as in Eq.~\eqref{eq:eta}. In the following, we compute the anisotropic stress using the profile models to account for possible systematics arising from the parametrization of the total cluster mass. For our reference analysis, we adopt the NFW model, which has been shown to provide the best fit of the stacked lensing signal from the analysis of Ref.~\cite{Umetsu16}.
%
 \section{Dataset and mass profiles}\label{sec:data}
%
    Our target is a sample of nine massive galaxy clusters, namely: Abell 383 (A383), Abell 209 (A209), RX J2129.7$+$0005 (R2129), MS2137$-$2353 (MS2137), RXC J2248.7$-$4431 (R2248, also named Abell S1063), MACS J1931.8$-$2635 (M1931), MACS J1115.9$+$0129 (M1115), MACS J1206.2$-$0847 (M1206), and MACS J0329.7$-$0211 (M329), spanning the redshift range $0.18<z\leq0.45$. The sample has been extensively studied within the CLASH (\textit {e.g.} Ref.~\cite{Postman12}) and CLASH-VLT (\textit{e.g.} Ref.~\cite{Rosati14}) collaborations. The full CLASH sample comprises twenty-five galaxy clusters, of which twenty were selected for their high X-ray temperatures ($kT_{\mathrm{X}} > 5$~keV) and morphologically regular X-ray appearance, as observed with \textit{Chandra}. The two-dimensional weak-lensing analysis of \cite{Umetsu18} clearly detected the elliptical shapes of cluster halos at a significance of $5\sigma$ in a sample of 20 CLASH clusters. However, the measured ellipticity for the CLASH sample shows a mild $1\sigma$ hint of rounder halo shapes relative to the $\Lambda$CDM prediction.
    The degree of dynamical relaxation in this sample - a fundamental assumption for the analysis presented here - will be further examined below.
    
    For our nine clusters, deep wide-field multi-band imaging was obtained and analyzed as part of the CLASH program, as presented in Ref.~\cite{Umetsu_2014_CLASHmasses}. For eight clusters (excluding R2248), the multi-band photometry was acquired with Suprime-Cam (\textit{e.g.} Ref.~\cite{Miyazaki_2002_SuprimeCam}) on the 8.2\,m Subaru Telescope. For R2248, the imaging was instead obtained with the Wide-Field Imager on the 2.2\,m MPG/ESO telescope at La Silla, as reported by Ref.~\cite{Gruen_2013_AS1063wl}. We refer the reader to Ref.~\cite{Umetsu_2014_CLASHmasses} for full details.
    
    As for the spectroscopic information, essential for the kinematic analysis, the main source of spectroscopic data for the nine clusters is the CLASH-VLT program (\textit{e.g.} Ref.~\citep{Rosati_2014_CLASH}), based on observations taken with the VIsual Multi-Object Spectrograph (VIMOS; \textit{e.g.}, Ref.~\cite{LeFevre_2003_VIMOS}), and with the Multi Unit Spectroscopic Explorer (MUSE; \textit{e.g.} Ref.~\cite{Bacon_2010_MUSEinstrument}), at the Very Large Telescope. MUSE observations are available for eight of the nine clusters, with the exception of MS2137.   
%
    \subsection{Lensing mass profiles}
%
    In this work, we derive marginalized posterior distributions for the $(r_{-2}, r_{200})$ parameters of our nine CLASH clusters, using the combined weak and strong-lensing data products presented in Ref.~\cite{Umetsu16}, assuming the three mass profile models introduced in Section~2.1. To this end, we use the piecewise-defined convergence profiles and total covariance matrices obtained by Ref.~\cite{Umetsu16} for the CLASH clusters (see their Appendix~B). In the lensing analysis, the total covariance matrix accounts for multiple sources of uncertainty (see Ref.~\cite{Umetsu20}), including statistical measurement errors, systematic effects primarily due to the residual mass-sheet degeneracy, the cosmic noise covariance from uncorrelated large-scale structures projected along the line of sight, and intrinsic variations in the lensing signal at fixed halo mass arising from halo triaxiality and projection effects.

    For all models, we adopt uninformative, log-uniform priors on both the halo mass, $M_{200}$, and the halo concentration, $c_{200} \equiv r_{200}/r_{-2}$, with prior ranges defined as $M_{200} \in [10^{14}, 10^{16}]\,h^{-1}\,M_\odot$ and $c_{200} \in [1, 20]$. For each cluster and each mass model, we then infer the posterior probability distribution in the $(r_{-2}, r_{200})$ parameter space. The resulting parameter constraints are summarized in the third and fourth columns of Table~\ref{tab.clusters}.
%
    \subsection{Kinematic mass reconstruction}
%
    The kinematic mass profiles are derived by means of the \textsc{MG-MAMPOSSt} code (\textit{e.g.} Ref.~\cite{Pizzuti2021,Pizzuti:2022ynt}), a version of the \textsc{MAMPOSSt} method of Ref.~\cite{Mamon01} which jointly reconstructs the velocity anisotropy and the mass profiles of clusters\footnote{Note that in general, the algorithm can be applied to any auto-gravitating system of particles moving smoothly under the influence of a gravitational potential.} using data of projected position, $R$, and line-of-sight velocities, $v_z$, in the cluster rest-frame, for cluster members. The points $(R,v_z)$ define the projected phase-space (p.p.s. hereafter).
    
    In this work, we employ the same setup as in
    Ref.~\cite{Biviano25Anis}; in particular, we consider the members selected in there by the application of the CLUster Membership in Phase Space (\textsc{CLUMPS}) method of Ref.~\cite{Biviano21}, further refined by Ref.~\cite{Biviano25Anis}.

    For each cluster, we consider galaxies up to a projected radius $R = r_{200}^L$, where $ r_{200}^L$ is the value of the virial radius estimated by the lensing analysis of Ref.~\cite{Umetsu18}. This choice limits the \textsc{MG-MAMPOSSt} analysis to a region in which the Jeans equation is more reliably applied, given that external cluster regions are less likely to be in dynamical equilibrium.
    
    As we have done in previous works (see \textit{e.g.} Ref.~\cite{Pizzuti2025a,Pizzuti2025b}), we checked that a variation of $\sim 10\%$ of this limit does not produce relevant effects on the final results. As for the lower limit, we excluded the central region $R<0.05 \,\text{Mpc}$ where the total mass profile and the dynamics is strongly influenced by 
    the brightest cluster galaxy (BCG) that dominates the gravitational potential (\textit{e.g.} Ref.~\cite{Sartoris2020,Biviano:2023oyf}). 

    Assuming that the 3-dimensional velocity distribution of member galaxies is Gaussian, the \textsc{MG-MAMPOSSt} procedure solves the Jeans equation to obtain the radial velocity dispersion $\sigma_r^2(r)$ for a set of parameters describing the mass profile $M(r)$, the velocity anisotropy profile, $\beta(r)$, and the number density $\nu(r)$. The radial dispersion is then projected along the line-of-sight. The code computes the probability $q(R,v_z|\vec{\theta})$ of finding a galaxy at the point $(R,v_z)$ of the phase space, given the set of parameters $\vec{\theta}$. The \textsc{MG-MAMPOSSt} log likelihood is then given by
        \begin{equation} \label{eq:mamlike}
         \ln \mathcal{L}_{MAM}(\vec{\theta}) = \sum_{i=1}^N \ln \Big[q\big(R^{(i)},v_z^{(i)}|\vec{\theta}\,\big)\Big],
        \end{equation}
    where the sum runs over the $N$ galaxies in the p.p.s.

    For the Jeans analysis, we consider the generalized Tiret model for the velocity anisotropy profile (\textit{e.g.} Ref.~\cite{Mamon19,Biviano24}),
        \begin{equation}
         \label{eq:BP}
         \beta_{gT}(r) = \beta_0 + (\beta_\infty - \beta_0)\frac{r}{r+r_\beta}\,,
        \end{equation}
    where $\beta_0$ and $\beta_\infty$ represent the anisotropy at $r=0$ and at very large radii, $r \gg r_{\beta}$, respectively. $r_\beta$ is a scale radius which is usually set to be equal to $r_{-2}$ of the cluster mass profile. In this work, we also explore the possibility of keeping $r_\beta$ as an additional free parameter; we further consider a different model of velocity anisotropy, the Pizzuti \& Biviano profile, or exponential Tiret (BP hereafter, see Maraboli et al., in prep.),

        \begin{equation}\label{eq:BP}
         \beta_{BP}(r) = \beta_0 \left(1 - \frac{r}{r+r_\beta}\right) +\beta_\infty \left[\frac{r}{r+r_\beta}+\frac{r^2}{r_\beta^2}e^{-\big(\frac{r}{r_\beta}\big)^2}\right]\,,
        \end{equation}
    which allows for a bump in the velocity anisotropy at $r\sim r_\beta$.
    
    Both gT and BP models are general enough to capture quite a broad range of orbit phenomenology for galaxies in clusters.\textsc{MG-MAMPOSSt} works using the scaled anisotropy parameters, $\mathcal{A}_0,\mathcal{A}_\infty$, defined as $\mathcal{A}_{0/\infty} = (1-\beta_{0/\infty})^{-1/2}$. Radial (tangential) orbits correspond to $\mathcal{A} > 1$ ($<1$, respectively), and $\mathcal{A} = 1$ corresponds to isotropic orbits ($\beta=0$). 
        \begin{figure}
         \centering
         \includegraphics[width=\textwidth]{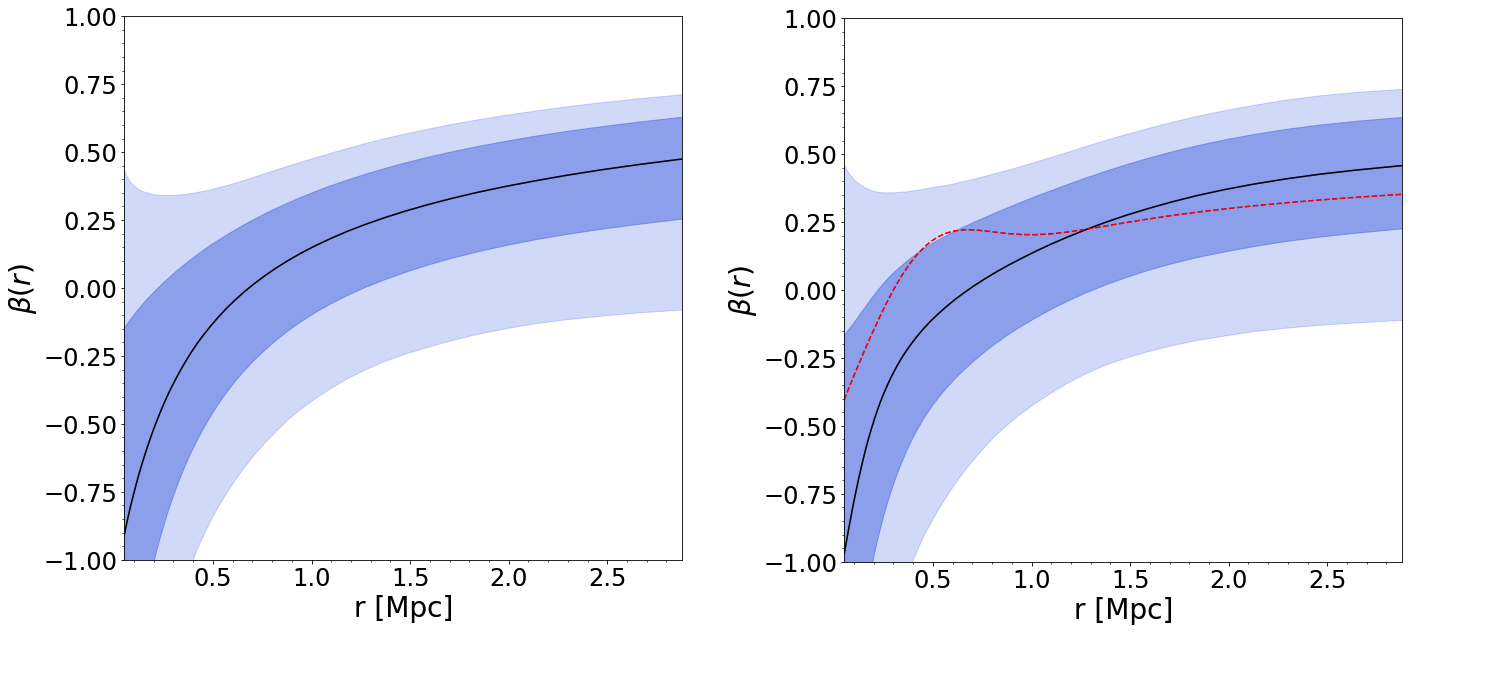}
         \caption{\label{fig:a209Anis} Anisotropy profiles of the cluster A209 reconstructed from the kinematic analysis with the \textsc{MG-MAMPOSSt} method Left: gT model. Right: BP model. In both cases, a NFW mass profile has been used. The darker and lighter shaded regions indicate $1$ and $2\,\sigma$ limits, respectively. The red dashed line in the plot on the right shows a BP profile with $\beta_0 =-0.5$, $\beta_\infty =0.5$ and $r_\beta = 0.5 \,\text{Mpc}$, for the only illustrative purpose of highlighting the features of the PB model.}
        \end{figure}

    In Figure \ref{fig:a209Anis} we plot an exemplary case of the anisotropy profiles. This corresponds to the A209 cluster obtained by means of the \textsc{MG-MAMPOSSt} procedure, assuming a gT model (left) and the BP model (right). The resulting profiles are almost identical within the errors. We also plot a BP model with $\beta_0 =-0.5$, $\beta_\infty =0.5$, and $r_\beta = 0.5 \,\text{Mpc}$ (red dashed line in the right plot) to highlight the features of this new model. The similarity of the gT and PB profiles indicates that the additional feature in the PB model with respect to the gT model does not provide an improved representation of the cluster $\beta(r)$.

    The last ingredient needed for the kinematic analysis is the number density profile of the galaxies. This can be obtained by fitting the radial density distribution $N(R)$ in projection, after correcting for the completeness of the sample, before running the \textsc{MG-MAMPOSSt} procedure. Note that only the scale radius $r_\nu$ of the profile is required in the Jeans analysis, as the normalization simplifies in the expression of $\sigma^2_r(r)$. Here, we rely on the results of Ref.~\cite{Biviano25Anis}, where the projected NFW (pNFW) model (\textit{e.g.} Ref. \cite{Bartelmann96}) and the King profile (\textit{e.g.} Ref.~\cite{King62}) are fitted to the p.p.s. with the method of Ref.~\cite{Sarazin80}, which doesn't require binning of the data. The pNFW model provides a better fit than the King model for all clusters except for the A209; the constraints on $r_\nu$ with the $1\,\sigma$ uncertainties are listed in column~7 of Table \ref{tab.clusters}. 
    
    To account for variability in the number density scale radius, $r_\nu$ is treated as a free parameter in \textsc{MG-MAMPOSSt}, with a Gaussian prior whose standard deviation corresponds to the 68\% confidence interval from the external fit.

    For each cluster, we perform a Monte-Carlo Markov-Chain (MCMC) sampling of the likelihood distribution of \eqref{eq:mamlike} for 110~000 points in the parameter space. We discard the first 10~000 points as a burn-in phase. As done in previous works (e.g., Ref.~\cite{Pizzuti25b}), the convergence is ensured by running 5 chains for every model choice and checking that the condition for the Geller-Rubin coefficient $\hat{R} <1$ is satisfied.

    We consider un-informative, flat priors in $r_{200} \in [0.3,5.0] \,\text{Mpc}$, $r_{s}, r_\beta \in [0.05,4.0] \,\text{Mpc}$, $\mathcal{A}_{0/\infty}\in [0.5,7.0]$; for $r_\nu$ the above mentioned Gaussian prior is assumed. The resulting $1\,\sigma$ and $2\,\sigma$ posterior distributions in the $r_{200}$-$r_{-2}$ space are shown for each cluster in Appendix \ref{app:kinVslens}, adopting the reference model gT for $\beta(r)$, compared to the lensing posteriors of Ref.~\cite{Umetsu16}.
    
    A visual inspection of Figures \ref{fig:lvsd_nfw}, \ref{fig:lvsd_Bur}, \ref{fig:lvsd_Her}, reveals that the kinematic reconstruction exhibits little variation when changing the mass model in the \textsc{MG-MAMPOSSt} procedure, differently from the lensing case. This confirms the robustness of the kinematic analysis against different choices of the total mass profile. This is further indicated by the fact that the value of $\Delta \chi^2 = -2 \Delta (\ln \mathcal{L}_{MAM})$ among the three models is always smaller than 1.0. An exemption occurs for the R2129 for which the NFW profile is mildly preferred with a  $\Delta \chi^2 \gtrsim 3$. 

    However, some clusters show a mild ($\lesssim 2 \sigma$ ) tension between the lensing and kinematic distributions, as one can visually determine again from Figures \ref{fig:lvsd_nfw}, \ref{fig:lvsd_Bur} and \ref{fig:lvsd_Her}; and by comparing the values listed in columns 3 and 4 of Table \ref{tab.clusters} with the corresponding values in columns 5 and 6, where we report the \textsc{MG-MAMPOSSt} constraints on $r_{200}$ and $r_s$ for the NFW+gT case.
    
    The shift is  evident in the scale radius of A209 and in the virial radius of M1115, where the kinematic analysis predicts systematically lower values of $r_s$ and $r_{200}$, respectively, with respect to the lensing case. The opposite trend is instead found for R2248, with overall larger dynamical masses. While M1115 does not exhibit strong evidence of unrelaxation (\textit{e.g.} Ref.~\cite{Donahue16}), A209 may have undergone (or is currently undergoing) a merging event, see Ref.~\cite{Jiménez-Teja_2018} and references therein. As for R2248, an indication of an off-axis merging event has been found by Ref.~\cite{Mercurio_2021}, suggesting that the cluster is not dynamically relaxed. We marked in bold these clusters in Table~\ref{tab.clusters}.

    It is also worth pointing out the slight tension in M1206, which was not found in the case study of Ref.~\cite{Pizzuti_2016}. In this, full consistency was shown between the kinematic and lensing profiles. As explained in Ref. \cite{Biviano:2023oyf}, the new p.p.s. membership selection, based on the CLASH-VLT+MUSE data, has been able to identify a foreground group which is located close to $R=0$ in projection, and which therefore is not considered in the \textsc{MG-MAMPOSSt} run. However, this substructure has not been excluded in the strong+weak lensing analysis of Ref.~\cite{Umetsu16}, producing a slight disagreement between the two mass determinations.
        \begin{table}
         \centering
         \small
            \begin{tabular}{cccccccccc} 
             \toprule
                Cluster & $z$ & $r_{200}^{L}\,\, $  & $r_{s}^{L}\,\, $  & $r_{200}^{D}\,\, $	 & $r_{s}^{D}\,\, $ & $r_{\nu}\,\, $ & $N_{200}$  & $A^2$ & $f_{sub}$\\[0.2cm]
             \midrule
                A383 & 0.187 & $1.78^{+0.43 }_{-0.41}$ & $0.33^{+0.30 }_{-0.24}$  & $1.79^{+0.15 }_{-0.14}$ & $0.23^{+0.24 }_{-0.19}$  & $0.47^{+0.08}_{-0.07} $& 425 & 0.65 & 0.20 \\[0.2cm]

                \textbf{A209} & \textbf{0.209} & {\boldmath $2.33^{+0.39 }_{-0.41}$} & {\boldmath $1.02^{+0.66 }_{-0.56}$} & {\boldmath $2.39^{+0.17 }_{-0.16}$ } & {\boldmath $0.61^{+0.54 }_{-0.43}$} & {\boldmath$0.66^{+0.03}_{-0.04} $ } & \textbf{864} & \textbf{0.75} & \textbf{0.40}\\[0.2cm]

                R2129 & 0.234 & $1.62^{+0.32 }_{-0.31}$ & $0.31^{+0.25 }_{-0.20}$ & $1.48^{+0.44 }_{-0.54}$ & $1.46^{+2.13 }_{-1.51}$ & $1.02^{-0.21}_{+0.26}$ & 185 & 0.36 & 0.46 \\[0.2cm]

                MS2137& 0.313 & $2.02^{+0.50 }_{-0.56}$ & $0.83^{+0.92 }_{-0.72}$ & $1.49^{+0.38 }_{-0.50}$ & $0.91^{+2.05 }_{-0.97}$  & $1.09_{-0.25}^{+0.33}$& 140 & 0.75 & 0.47 \\[0.2cm]

                \textbf{R2248} & \textbf{0.348} & {\boldmath$2.25^{+0.54 }_{-0.51}$} & {\boldmath$0.72^{+0.77 }_{-0.55}$} & {\boldmath$2.52^{+0.24 }_{-0.22}$} & {\boldmath$0.72^{+0.62 }_{-0.49}$} & {\boldmath$0.97^{+0.10 }_{-0.09}$} & \textbf{688} & \textbf{1.02} & \textbf{0.50} \\[0.2cm]

                M1931 & 0.352 & $2.27^{+0.67 }_{-0.66}$ & $0.78^{+1.03 }_{-0.69}$ & $1.93 \pm 0.30 $ & $0.64^{+1.16 }_{-0.66}$ & $1.13^{+0.29}_{-0.23}$ & 218 & 0.20 & 0.10 \\[0.2cm]

                \textbf{M1115} & \textbf{0.355} &  {\boldmath $2.16^{+0.31 }_{-0.34}$} & {\boldmath $0.77^{+0.48 }_{-0.41}$} & {\boldmath $1.67^{+0.32 }_{-0.36}$} & {\boldmath $1.36^{+1.69 }_{-1.16}$} & {\boldmath $1.06^{+0.18 }_{-0.14}$}  & \textbf{ 363 } & \textbf{0.67} & \textbf{0.31} \\[0.2cm]

                M1206 & 0.439 & $2.15^{+0.32 }_{-0.34}$ & $0.62^{+0.45 }_{-0.37}$ & $2.02^{+0.16 }_{-0.15}$ & $0.28^{+0.33 }_{-0.23}$ & $ 0.67^{+0.10}_{-0.12}$ & 409 & 0.49 & 0.22 \\[0.2cm]

                M329 & 0.450 & $1.66^{+0.26 }_{-0.27}$ & $0.25^{+0.18 }_{-0.15}$  & $1.75^{+0.29 }_{-0.35}$ & $0.67^{+1.51 }_{-0.68}$  & $0.70^{+0.15}_{-0.13} $ & 199 & 0.40 & 0.0 \\[0.2cm]

                \bottomrule 
            \end{tabular}
         \caption{\label{tab.clusters} Summary of the constraints on $r_{200}$ and $r_s$ along with their 2$\sigma$ uncertainties for the set of nine CLASH clusters analysed here. All radii are in Mpc units. We adopt the reference model NFW for both lensing and kinematic masses, with a gT model for the velocity anisotropy profile. First column: cluster denomination. Second column: redshift. Third and fourth columns: constraints from the lensing analysis of \cite{Umetsu16}. Columns five and six: constraints from \textsc{MG-MAMPOSSt}. Column seven: scale radius of the number density profile along with its 1$\sigma$ uncertainties, adopted as Gaussian priors in the kinematic MCMC run. Column eight: number of selected members within $r_{200}^L$. Column nine: Anderson-Darling coefficient. Column ten: fraction of galaxies in substructures. We marked in bold the clusters with signs of an un-relaxed state.}
        \end{table}
%

%
 \section{Results}\label{sec:results}
%
    We reconstruct $\eta(r)$ by applying Eq.~\eqref{eq:eta} as follows: we consider $10^5$ arrays of parameters ($r^L_{200}$, $r^L_{s}$, $r^D_{200}$, $r^D_{s}$), randomly extracted from the MCMC samples of the lensing and kinematic posterior distributions. We then compute $\eta(r)$ in the radial range $[\,0.1\,\text{Mpc},\,1.2\, r_{200}^{L}\,]$, where $r_{200}^{L}$ is the best fit value of $r_{200}$ found by the lensing analysis of Ref.~\cite{Umetsu16}. The lower limit is chosen to exclude the central cluster region, where the uncertainties on the lensing model become larger (\textit{e.g.} Ref. \cite{Zitrin15,Umetsu16}) and the dynamics becomes more influenced by the BCG.


    As for the upper limit, note that extending the analysis beyond $r_{200}$ is generally not appropriate, as the assumption of dynamical equilibrium may break down at such large radii. For the lensing data, parameter inference is limited to $r = 2\,h^{-1}\,\mathrm{Mpc} \simeq 2.9\,\mathrm{Mpc}$, beyond which the contribution from surrounding large-scale structure becomes increasingly significant (see Figure 3 of Ref.~\cite{Umetsu16}). 
    Keeping all of this in mind, we quoted as our reference value for each cluster $\eta(r=1.0\,\text{Mpc})$ as a fair trade-off between not losing too much information in the central region and avoiding the innermost core. Note also that around $r\sim1 \,\text{Mpc}$ the radial profiles $\eta(r)$ tend to flatten (see Figure~\ref{fig:radial} of Appendix~\ref{app:radial}). Moreover, for many viable MG models, the innermost cluster region is totally screened, masking additional effects induced by departures from GR (see \emph{e.g.} Ref. \cite{Pizzuti_2024b}). 

\begin{figure}
     \centering
     \includegraphics[width=\textwidth]{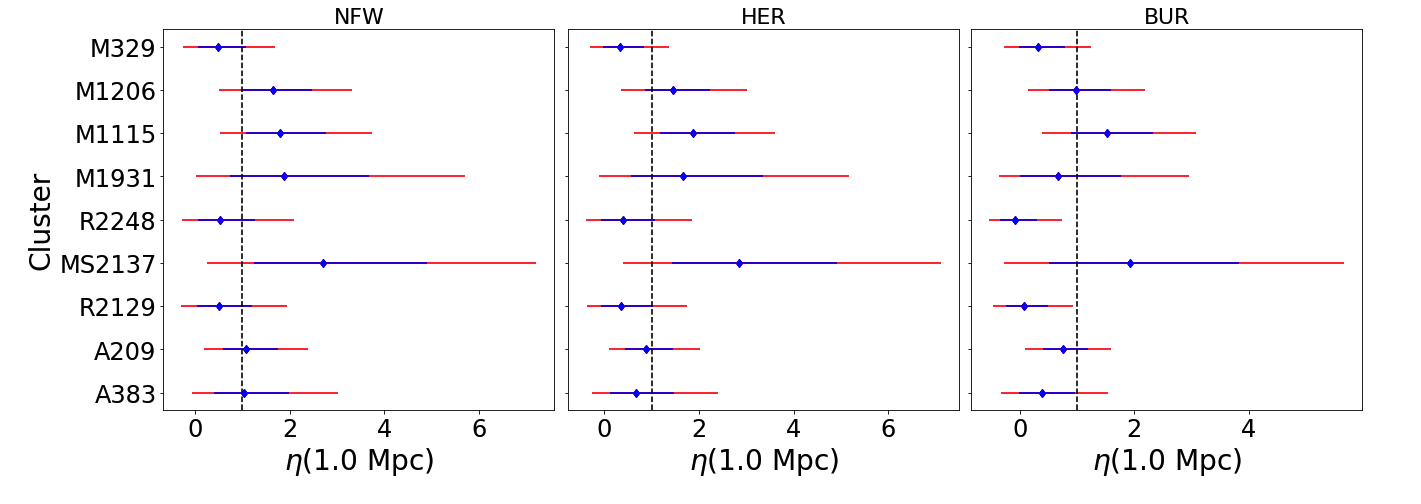}
     \caption{\label{fig:eta} Constraints on $\eta(r=1.0\, \text{Mpc})$ for the nine clusters in the sample. In each plot, a gT model is assumed for the velocity anisotropy profile in the kinematic analysis. The blue segments refer to $68\%$ C.L., while the red lines indicate the $95\%$ C.L. For Left: NFW model. Middle: Hernquist model. Right: Burkert model. The black dashed vertical lines mark the GR expectation $\eta = 1$.}
\end{figure}

    In Figure \ref{fig:eta} we plot the constraints of $\eta(r=1.0\, \text{Mpc})$ derived at 1$\sigma$ and 2$\sigma$ for all the clusters in the sample. Each plot corresponds to a different model of $M(r)$, while the gT profile is adopted for the velocity anisotropy in all cases. The full radial profiles of $\eta(r)$ for the reference model NFW+gT are also reported in Figure~\ref{fig:radial} of Appendix~\ref{app:radial}. 

    Independently of the mass model assumed, the GR expectation $\eta=1$ is always included within 2$\sigma$, except for the Burkert case, where two clusters (R2248 and R2129) exhibit a $\sim 3 \sigma$ tension in favour of $\eta <1$. Note, however, that the lensing analysis disfavors the Burkert model when compared to the other two profiles (see \textit{e.g.} Ref. \cite{Umetsu16}). The individual constraints are also shown in the second, fourth, and sixth columns of Table \ref{tab:eta}, for NFW, Hernquist, and Burkert total mass profiles, respectively. 

\begin{figure}
     \centering
     \includegraphics[width=\textwidth]{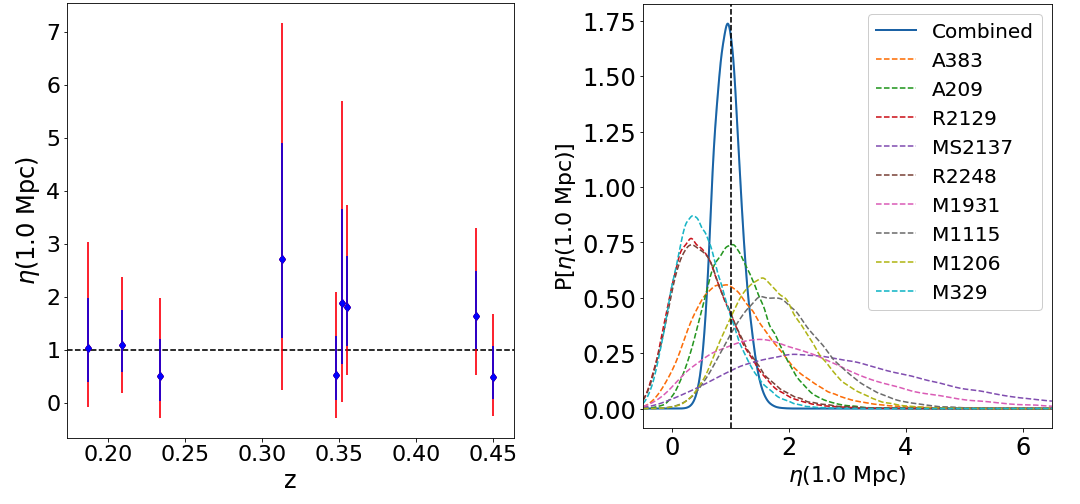}
     \caption{\label{fig:redshift} Left: constraints on $\eta(r=1.0\, \text{Mpc})$ as a function of the redshift for the NFW+gT model. The color code is the same as in Figure~\ref{fig:eta}. Right: Posterior distributions from the analysis of each single cluster (dashed lines), computed at $r= 1.0\,\text{Mpc}$, and combined distribution (solid line).}
    \end{figure}

\begin{figure}
     \centering
     \includegraphics[width=\textwidth]{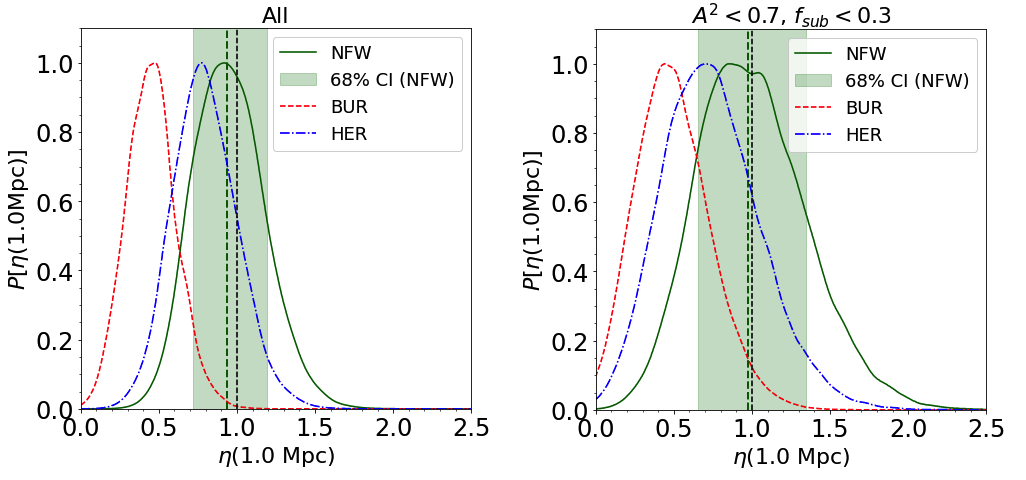}
     \caption{\label{fig:etacombined} Left: posterior distributions of $\eta(r= 1.0\,\text{Mpc})$ for the three mass models adopted in our analysis. The gT profile is assumed for the velocity anisotropy. Right: same, but considering only clusters with $A^2 < 0.6$ and  $f_{sub} < 0.3$. In both plots, the dashed green vertical lines correspond to the median of the NFW-case distribution, and the green shaded regions indicate the corresponding $1\,\sigma$ confidence interval. The  black short-dashed vertical lines indicate the "GR" value $\eta=1$}
    \end{figure}

    In the left panel of Figure \ref{fig:redshift}, we plot $\eta(r=1.0 \, \text{Mpc})$ as a function of the cluster redshift $z$ for the reference model NFW+gT. No evidence of time evolution has emerged from our analysis in the redshift range explored; this statement motivates the possibility of combining the single marginalized posterior distributions $P[\eta(r=1\,\text{Mpc})]$ to get the joint constraint on the anisotropic stress for our cluster sample. The distributions for all the clusters, as well as the combined posterior for the NFW+gT case, are plotted in the right panel of Figure~\ref{fig:redshift}.
    The left panel of Figure~\ref{fig:etacombined} shows the effect on the distribution of adopting different mass models in the analysis. While NFW and Hernquist profiles show similar results, in agreement with GR expectation $\eta = 1$ within 68\% C.L. and 95\% C.L., respectively, the Burkert ansatz provides a posterior significantly shifted towards $\eta < 1$. We find $\eta(r=1.0\,\text{Mpc}) = 0.93^{+0.22}_{-0.18}\, (1\sigma)\, ^{+0.48}_{-0.40} \,(2\sigma) \pm 0.47 \,(\text{syst})$, where the systematic uncertainties reflect the variation of the peak of the distribution induced by the choice of the mass profile.
%

    The $\sim 3 \sigma$ tension observed in the Burkert model can be explained again by inspecting the distributions $P(r_{200},r_{-2})$ in Figure  \ref{fig:lvsd_Bur}. Compared to the NFW and Hernquist cases, the lensing analysis for the Burkert profile prefers overall lower values of $r_{-2}$ and $r_{200}$ for many clusters, with tighter constraints. From Figure 4 of \cite{Umetsu16}, it is clearly seen that when the Burkert profile is forced to fit the stacked lensing signal, it tends to favor the inner region where the signal-to-noise ratio is higher. As a consequence, the outer density profile is significantly underestimated. A similar bias may be present in individual cluster analyses, potentially explaining why both $r_{200}$ and $r_{-2}$ tend to be underestimated when modeling the lensing signal with a Burkert profile.


%
    \begin{table}[h!]
     \centering
    \renewcommand{\arraystretch}{1.2} 
        \begin{tabular}{c ccc} 
         \toprule
         Cluster & NFW & Her & Bur \\
         \midrule
         A383 & $1.04^{+1.99 }_{-1.11}$ & $0.66^{+1.75 }_{-0.92}$ & $0.39^{+1.14 }_{-0.72}$\\[0.2cm]
         A209 & $1.09^{+1.29 }_{-0.91}$ & $0.87^{+1.14 }_{-0.78}$ & $0.76^{+0.84 }_{-0.66}$  \\[0.2cm]
         R2129 & $0.51^{+1.45 }_{-0.80}$ & $0.36^{+1.41 }_{-0.74}$ & $0.07^{+0.86 }_{-0.54}$ \\[0.2cm]
         MS2137 & $2.72^{+4.48 }_{-2.46}$ & $2.85^{+4.26 }_{-2.45}$ & $1.92^{+3.75 }_{-2.19}$ \\[0.2cm]
         R2248 & $0.54^{+1.56 }_{-0.82}$ & $0.38^{+1.48 }_{-0.77}$ & $-0.08^{+0.82 }_{-0.45}$\\[0.2cm]

         M1931 & $1.89^{+3.82 }_{-1.86}$ & $1.67^{+3.50 }_{-1.79}$ & $0.67^{+2.29 }_{-1.03}$ \\[0.2cm]

         M1115  & $1.81^{+1.92 }_{-1.28}$ & $1.87^{+1.73 }_{-1.25}$ & $1.53^{+1.56 }_{-1.14}$\\[0.2cm]

         M1206 & $1.65^{+1.66 }_{-1.13}$ & $1.45^{+1.57 }_{-1.10}$ & $0.98^{+1.21 }_{-0.83}$ \\[0.2cm]
         M329 & $0.49^{+1.20 }_{-0.74}$ & $0.33^{+1.04 }_{-0.64}$ & $0.33^{+0.92 }_{-0.61}$ \\[0.2cm]
         \midrule
         Combined & $0.93  ^{+0.48} _{-0.40}$  &  $0.78^{+0.44 }_{-0.38}$  & $0.45^{+0.34 }_{-0.30}$ \\
         \bottomrule 
        \end{tabular}
     \caption{\label{tab:eta} Constraints at $2\,\sigma$ on $\eta (r=1.0 \,\text{Mpc})$ for each individual cluster, adopting the three mass models described in Section~\ref{sec:theo}, and for the combined distributions (last row).}
    \end{table}
%
    \subsection{Impact of dynamical relaxation and velocity anisotropy}
%
    In order to establish the robustness of our analysis, let us consider the effect of departures from dynamical relaxation and the presence of substructures on the final constraints on $\eta$. As done in previous works (\textit{e.g.} Ref.~\cite{Barrena24,Pizzuti2025b}), we adopt the Anderson-Darling coefficient $A^2$ of Ref.~\cite{Anderson52}, which quantifies deviation from Gaussianity of the line-of-sight velocity distribution, in order to estimate the magnitude of departures from dynamical equilibrium. The total fraction of galaxies in substructures, $f_{\rm sub}$, is further considered as a proxy for the disturbed state of clusters; for our sample, we use the values estimated in Ref.~\cite{Biviano25Anis}, by running the DS+ algorithm of Ref. \cite{Benavides23}, which combines the information of spatial distribution and line-of-sight velocity to detect subclusters. The values of $f_{\rm sub}$ and $A^2$ are listed in the last columns of Table \ref{tab.clusters}.

\begin{figure}
         \centering
         \includegraphics[width=\textwidth]{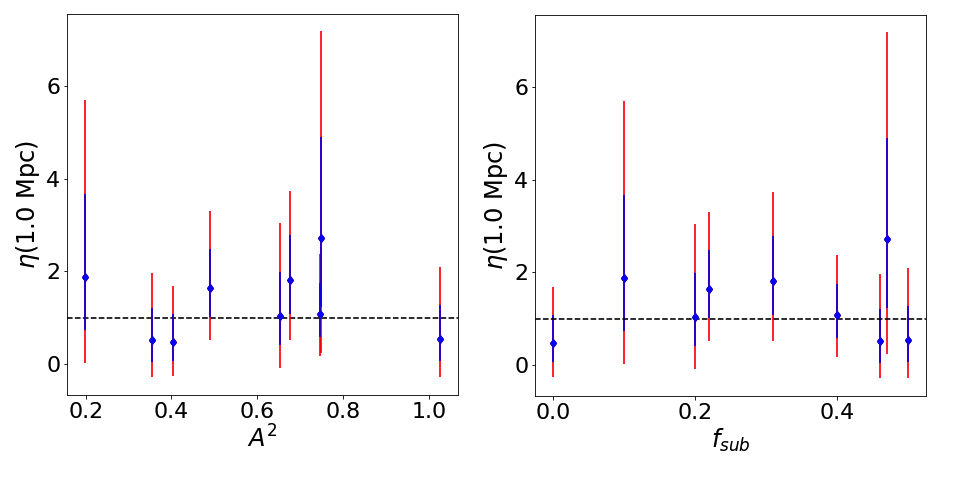}
         \caption{\label{fig:Asub} Constraints on $\eta(r= 1\,\text{Mpc})$ for the NFW+gT reference case as a function of the Anderson-Darling coefficient $A^2$ on the left, and the fraction of galaxies in substructures $f_{\rm sub}$ on the right. The color code is as in Figure \ref{fig:eta}.}
        \end{figure}

    In Figure \ref{fig:Asub}, the constraints on $\eta(r=1.0\, \text{Mpc})$ as a function of $A^2$ (left) and $f_{\rm sub}$ (right) are shown for the NFW+gT model. By visually inspecting both plots, no strong evidences of correlation arise; moreover, when considering only the clusters with $A^2 < 0.7$ and $f_{\rm sub} < 0.3$, namely M329, M1931, M1206 and A383, no substantial change in the combined distribution is found, as it can be seen in the left panel of Figure \ref{fig:etacombined}. 
    It is worth pointing out that here we investigate the effects related to the kinematics of member galaxies. We are not considering other possible sources of systematics in the lensing mass profiles. As an example, the aforementioned foreground group in M1206 can contaminate the strong-lensing analysis by boosting the projected mass profile in the cluster core. In contrast, it has been excluded from the kinematic phase-space selection. This foreground contamination may therefore account for the discrepancy between the kinematic and lensing-based constraints on $(r_{-2}, r_{200})$. As discussed in Ref.~\cite{Biviano:2023oyf}, adding the mass profile of the foreground group to the cluster mass profile obtained by \textsc{MG-MAMPOSSt} reproduces the lensing mass profile very well.

\begin{figure}
         \centering
         \includegraphics[width=0.6\textwidth]{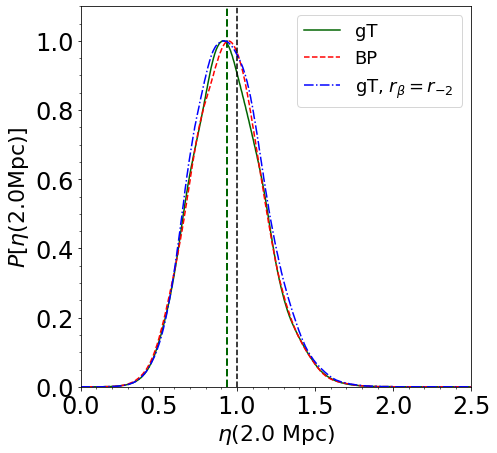}
         \caption{\label{fig:beta} Posterior of $\eta(r=1.0\,\text{Mpc})$ for gT model of $\beta(r)$ (green solid), gT model with $r_\beta = r_{-2}$ (blue dash-dotted), and BP model (red dashed). The green vertical line is the mode of the gT-case distribution. For all cases, the NFW mass model is assumed. The black short-dashed line is the GR expectation $\eta=1$. }
        \end{figure}

    As a last step, we discuss the impact of the velocity anisotropy profile model assumed in the kinematic analysis. Besides the gT model, we run \textsc{MG-MAMPOSSt} with the BP profile of Eq.~\eqref{eq:BP}; we also consider a third case where the usual assumption of $r_\beta = r_{-2}$ is adopted in the gT model. For each choice of the mass profile, we found that the resulting constraints are identical, independently of the model of $\beta(r)$, indicating that the choice of the model for the velocity anisotropy profile does not constitute a relevant source of systematics in this type of analysis. The posteriors obtained assuming different $\beta(r)$, for the reference NFW mass case, are further reported in Figure \ref{fig:beta}.
%

%
\section{Conclusions}\label{sec:conc}
%
    In this paper, we presented a parametric determination of the scale-dependent anisotropic stress $\eta(r) = \Psi(r)/\Phi(r)$ in the radial range $[0.1\,\text{Mpc},1.2\,r_{200}^L]$. The term $\eta (r)$ is determined from the combination of kinematic and lensing mass profiles of nine massive galaxy clusters of the CLASH/CLASH-VLT sample. We used high-quality photometric and spectroscopic data of positions and line-of-sight velocities of member galaxies to jointly reconstruct the mass and velocity anisotropy profiles by means of the \textsc{MG-MAMPOSSt} code of Ref. \cite{Pizzuti2021}. For different prescriptions for the mass profile models and the velocity anisotropy, we combined the kinematic mass with posteriors obtained by the strong+weak lensing analysis of Ref.~\cite{Umetsu16}, allowing the constraint of $\eta(r)$.
    
    For all clusters, when considering the Hernquist and NFW mass profiles, we found consistency with the expectation of GR ($\eta = 1$) within 2$\sigma$. For the Burkert model, two clusters show a quite strong ($\sim 3 \sigma$) indication of $\eta \ne 1$; however, the Burkert profile is disfavored by the strong+weak lensing fit of Ref.~\cite{Umetsu16}. 

    No evidence of redshift evolution has emerged from our analysis, which allowed us to combine the posterior distributions of the anisotropic stress for each cluster, obtaining the constraint $\eta(r=1.0\,\text{Mpc}) = 0.93^{+0.22}_{-0.18}\, (1\sigma)\, ^{+0.48}_{-0.40} \,(2\sigma) \pm 0.47 \,(\text{syst})$. The systematic encapsulates the effects of choosing different models for the lensing and kinematic mass profiles, while no variation induced by the choice of the anisotropy model has been found. 


    We further investigated possible relations between our results and the global degree of relaxation estimated from kinematical proxies, such as the Anderson-Darling coefficient, $A^2$, and the total fraction of galaxies in substructures $f_{\rm sub}$. No strong indication of correlation was found; when further restricting the analysis to relaxed clusters with $A^2 < 0.7$ and $f_{\rm sub} < 0.3$, the joint constraints do not show significant changes. 

    The present analysis underscores two main findings. First, adopting the NFW model yields bounds on $\eta$ that are approximately 40\% tighter than previous determinations in the same redshift range — obtained from clusters (\textit{e.g.} Ref.\cite{Pizzuti_2016}) -, and comparable to those from galaxy strong lensing (\textit{e.g.} Ref.~\cite{Guerrini24}) or from combined probes such as CMB, weak lensing, and galaxy clustering (\textit{e.g.} Ref.~\cite{Li2019}). The results are in agreement with theoretical predictions (\textit{e.g.} Ref.~\cite{Amendola:2014wma,Pizzuti19,Casas23}). This confirms the constraining power of galaxy clusters as a test for signatures of non-standard models, a relevant tool in view of the large amount of imaging and spectroscopic data that are becoming available in the present and future years. In particular, photometric surveys such as the Legacy Survey of Space and Time (LSST, \textit{e.g.} Ref.~\cite{LSST19}), the Vera C. Rubin Observatory, \emph{Euclid} (\textit{e.g.} Ref.~\cite{EuclidI}), and the Nancy Grace Roman Space Telescope (\textit{e.g.} Ref.~\cite{Wenzl_2022}) will provide deep and wide-field imaging, enabling high-precision weak lensing measurements and photometric identification of clusters over a large redshift range. At the same time, spectroscopic campaigns, such as DESI (\textit{e.g.} Ref.~\cite{DESI2016}) and 4MOST (\textit{e.g.} Ref.~\cite{4MOST2019}), will provide redshift and velocity measurements for tens of thousands of galaxies. These datasets will enable dynamical mass reconstruction and substructure analyses over large statistical samples of galaxy clusters. In parallel, dedicated cluster-focused surveys beyond CLASH, such as X-COP (\textit{e.g.} Ref. \cite{Eckert2017}), CHEX-MATE (\textit{e.g.} Ref. \cite{chexmate21}), and HeCS/HeCS-SZ (\textit{e.g.} Ref. \cite{Rines_2013,Rines_2016}), offer detailed mass profile determinations down to the cluster core by combining multi-wavelength data (X-ray, SZ, and optical), allowing stringent constraints on the dark sector and potential deviations from GR.

    On the other hand, the mismatch observed between lensing and kinematic mass profiles in some clusters of the sample underscores the necessity of a thorough assessment and calibration of systematic uncertainties. Furthermore, it points to the importance of identifying reliable observational diagnostics - such as $A^2$ or $f_{\rm gas}$ - that can serve as selection criteria for a "golden sample" of galaxy clusters, suitable for deriving robust constraints on alternative theories of gravity.

    Note that for a few systems in our sample, such as M1206 or R2248, the availability of kinematic data of the BCG as well as the precise measurements of gas and stellar components allowed a more detailed mass profile determination down to $\sim 1 $ kpc (\textit{e.g.} Ref.~\cite{Sartoris2020,Biviano:2023oyf}). At the same time, refined strong lensing analysis has been performed to obtain the mass distribution in the core of M1206 (\textit{e.g.} Ref. \cite{Bergamini23}). This upgraded information will be used in future work to refine the single-cluster bounds on $\eta(r)$ as well as to get new insights on the behavior of gravity between astrophysical and cosmological scales.



\acknowledgments

The authors acknowledge P. Rosati and the CLASH-VLT team for having provided the dataset used in this work. K.U. acknowledges support from the National Science and Technology Council of Taiwan (grant NSTC 112-2112-M001-027-MY3) and the Academia Sinica Investigator Award (grant AS-IA112-M04). A. M. Pombo is supported by the Czech Grant Agency (GA\^CR) project PreCOG (Grant No. 24-10780S).
\bibliographystyle{JHEP}
\bibliography{sample631}


\appendix
\section{Kinematic vs lensing Posteriors}
\label{app:kinVslens}
\begin{figure}
    \centering
    \includegraphics[width=.8\textwidth]{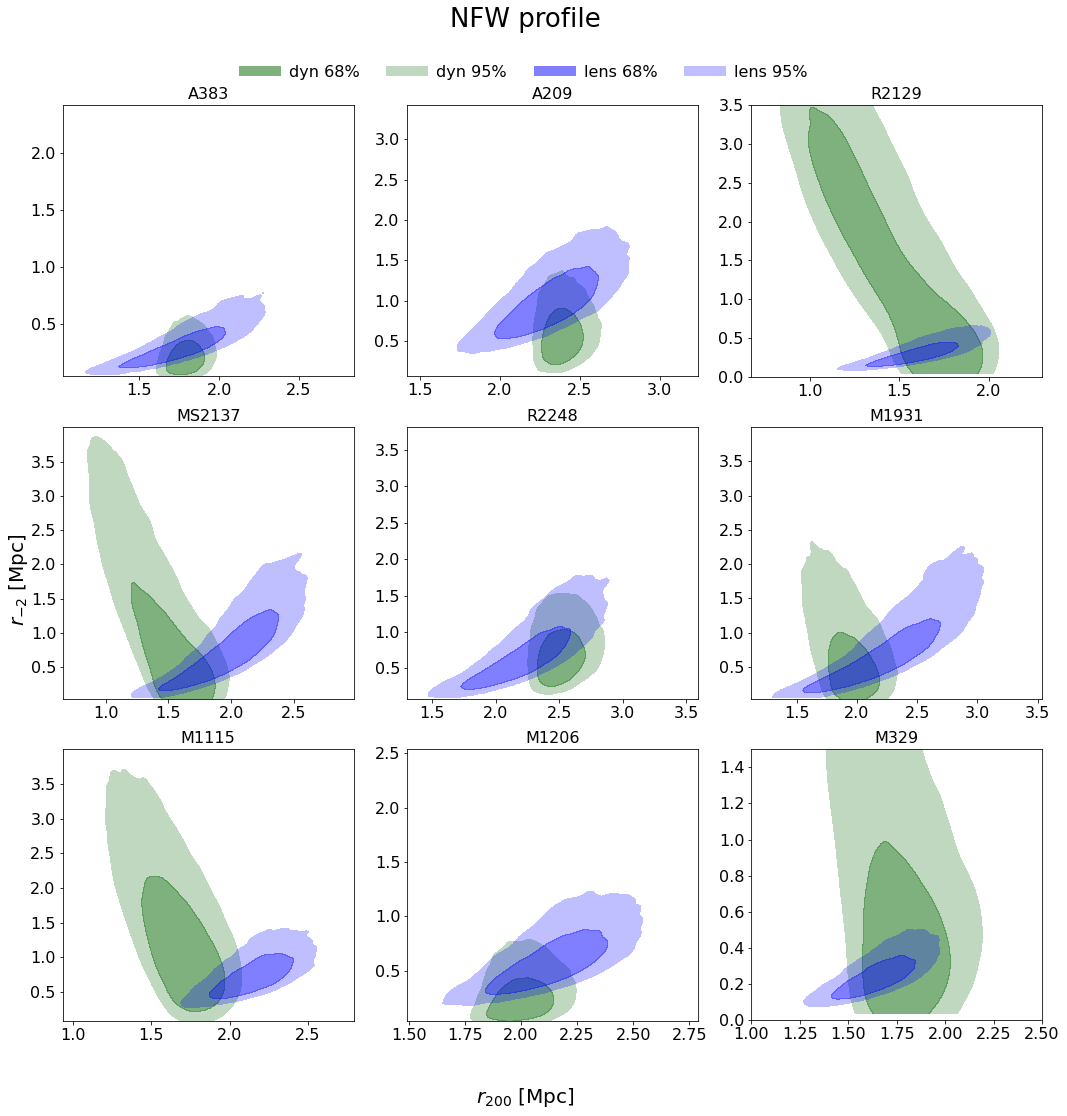}
    \caption{Two dimensional posterior distributions $P(r_{200},r_{-2})$ at 1 $\sigma$ (darker shaded regions) and 2 $\sigma$ (lighter shaded regions) for the lensing and kinematic analyses of the nine CLASH clusters. The total mass is described by a NFW profile, and we adopted a gT model for the velocity anisotropy in the \textsc{MG-MAMPOSSt} fit. Blue: lensing. Green: kinematics. \label{fig:lvsd_nfw}}
\end{figure}

\begin{figure}
    \centering
    \includegraphics[width=.8\textwidth]{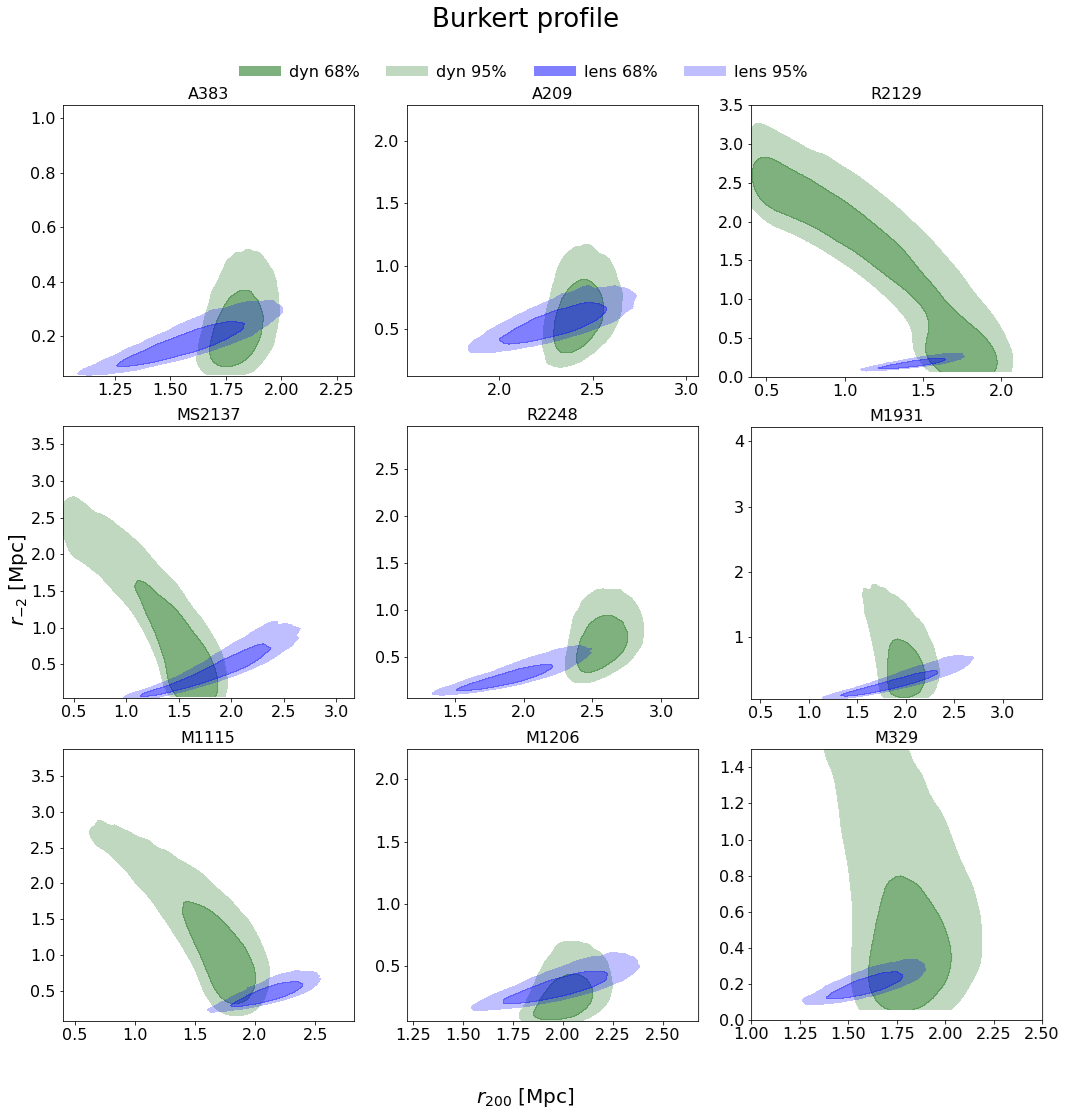}
    \caption{Same as Fig. \ref{fig:lvsd_nfw} but for the Burkert model assumed for the total mass profile. \label{fig:lvsd_Bur}}
\end{figure}

\begin{figure}
    \centering
    \includegraphics[width=.8\textwidth]{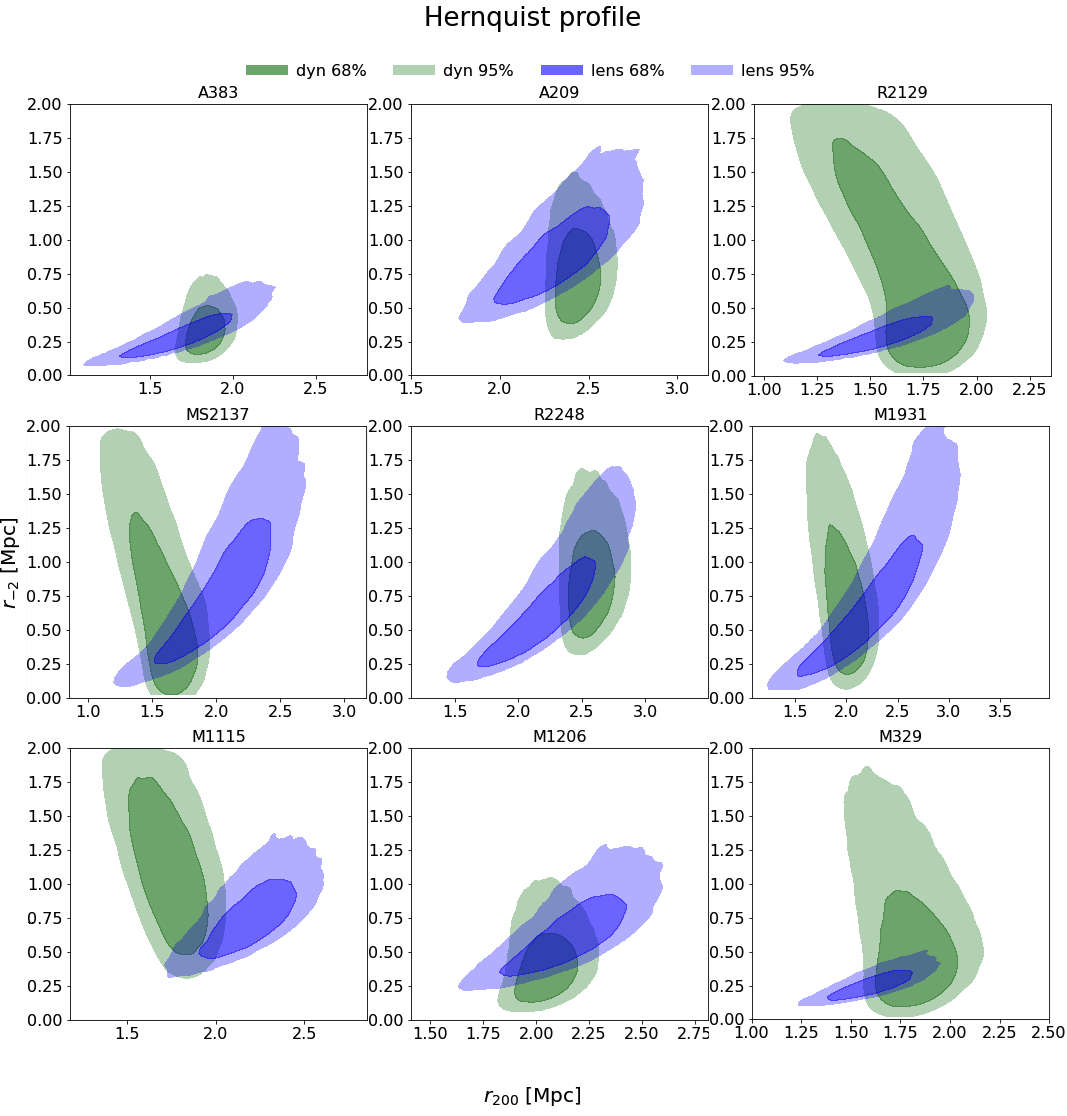}
    \caption{Same as Fig. \ref{fig:lvsd_nfw} but for the Hernquist model assumed for the total mass profile. \label{fig:lvsd_Her}}
\end{figure}

    In Figures  \ref{fig:lvsd_nfw}, \ref{fig:lvsd_Bur}, \ref{fig:lvsd_Her}  we present the 1 $\sigma$ and 2 $\sigma$ regions in the space $(r_{200}, r_{-2})$ for the lensing (blue) and \textsc{MG-MAMPOSSt} kinematic (green) analyses. Each Figure refers to a different choice of the model for the total cluster mass. Note that $r_{-2} = r_s/2 $ for the Hernquist profile and  $r_{-2} \simeq 3 r_s/2$ for the Burkert profile. 

    While the kinematic analysis shows consistency among the three mass choices, the lensing results appear very different for the Burkert case (which, as mentioned in Sec. \ref{sec:data}, has been found to poorly fit the stacked lensing signal in \cite{Umetsu16}). Clusters as A209, M1115, R2248 exhibit a mild ($\lesssim 2\sigma$) tension between the lensing and kinematic distributions, which translates to a constrain on $\eta \ne 1$. Interestingly, the disagreement is reduced for A209 when the Burkert model is assumed. This can follow from the fact that A209 is the only galaxy cluster in the sample showing a core in the galaxy distribution (Ref.~\cite{Biviano25Anis}). On the contrary, the tension increases significantly in the case of R2248 and R2129 for the Burkert model.


\begin{figure}
    \centering
    \includegraphics[width=.8\textwidth]{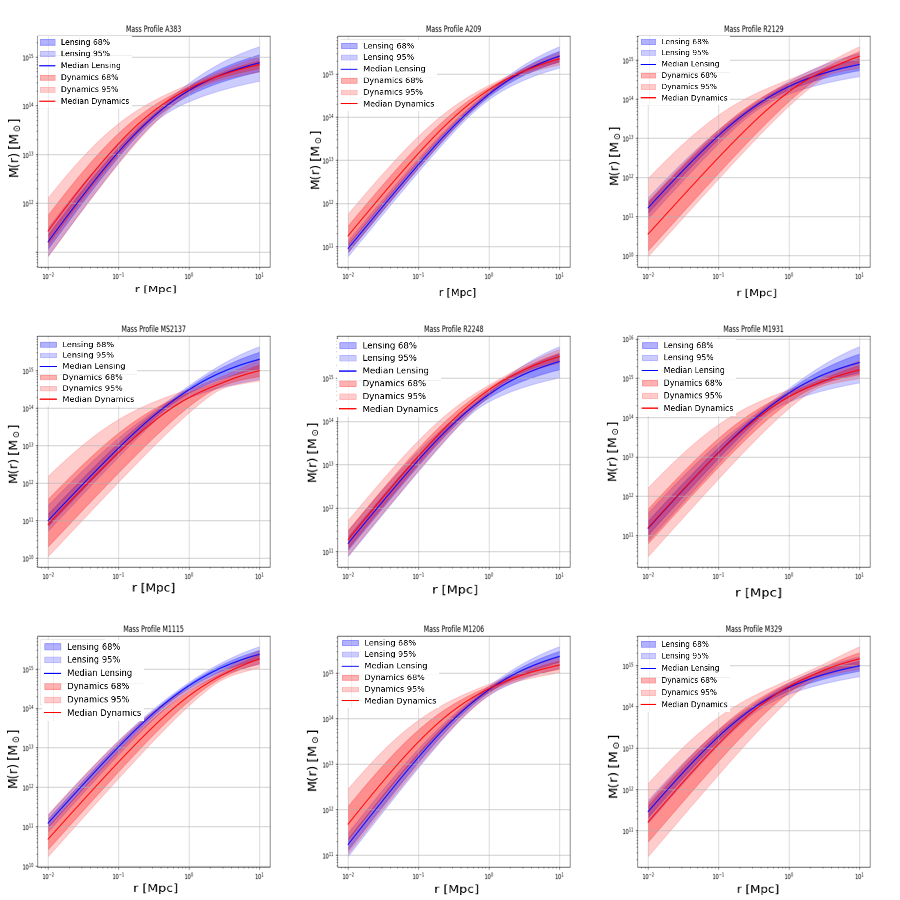}
    \caption{blue: strong+weak lensing radial mass profiles, in blue,  compared to the mass profiles derived with \textsc{MG-MAMPOSSt} in red. The darker and lighter shaded regions indicate the $1\sigma$ and $2\sigma$ interval, respectively. Solid lines are the median profiles\label{fig:profiles}. All profiles are parametrised with the NFW model.}
\end{figure}

Figure \ref{fig:profiles} further shows the radial mass profiles obtained from the lensing chains (blue solid line, blue shaded areas) and those resulting from the \text{MG-MAMPOSSt} pipeline (red). In all cases, we have assumed a NFW model. 

\section{Radial profiles of $\eta(r)$}
\label{app:radial}
\begin{figure}
    \centering
    \includegraphics[width=.8\textwidth]{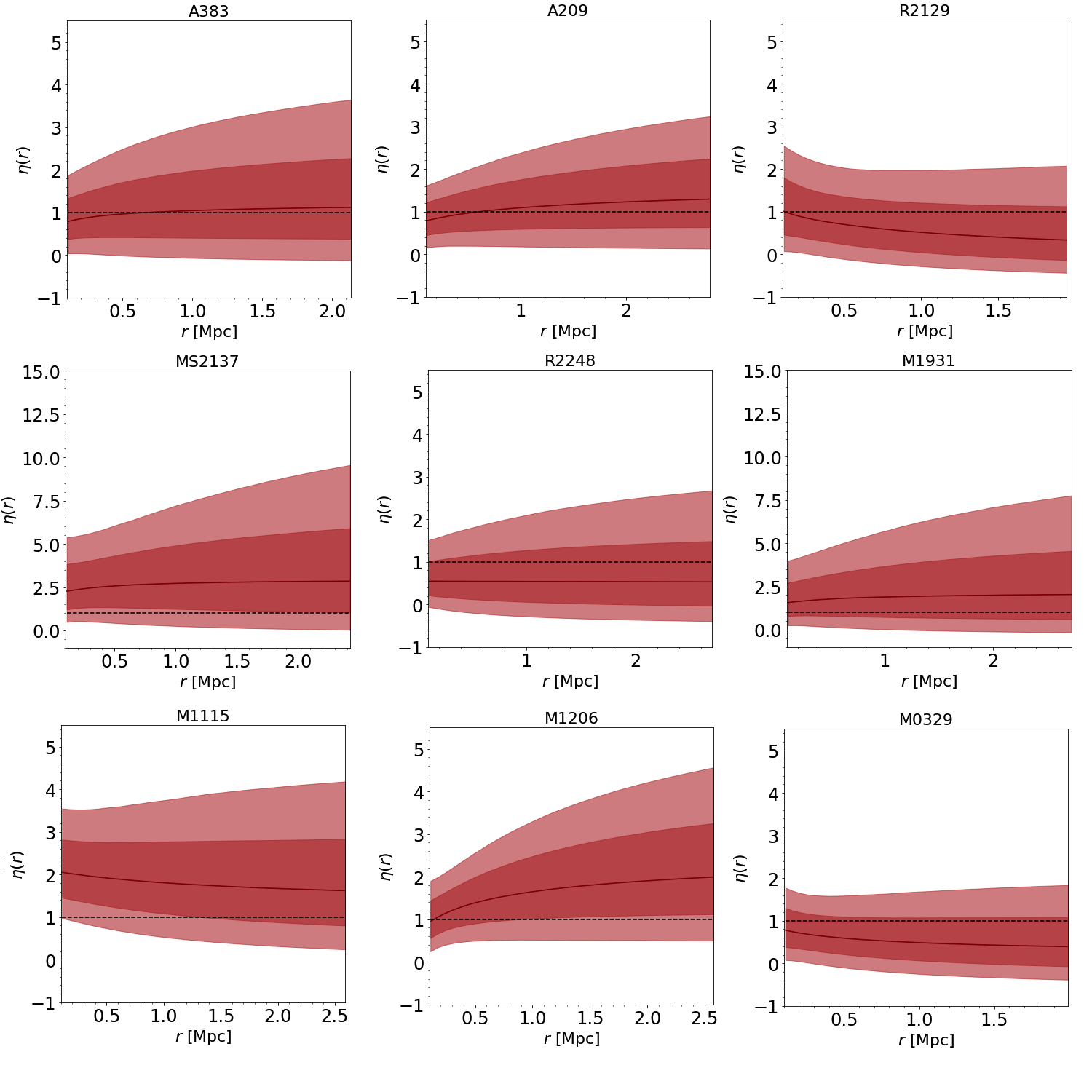}
    \caption{\label{fig:radial} Radial profiles $\eta(r)$ obtained for the nine clusters in the sample adopting the reference model NFW+gT. In each plot, shaded regions refer to  1~$\sigma$ (dark red) and 2~$\sigma$ (light red) intervals; the red solid line indicates the median. The GR expectation $\eta = 1$ is marked by the horizontal black dashed line.}
\end{figure}

    In Figure \ref{fig:radial} we plot the radial profiles of the anisotropic stress for the nine clusters in the sample, adopting a NFW+gT model. The darker and lighter shaded areas represent the 1~$\sigma$ and 2~$\sigma$ confidence region, respectively, while the median values are represented by the red solid line.






\end{document}